\documentclass[12pt]{article}

\begin{document}
\def \l {\Lambda}
\def \o {\Omega}
\newcommand{\lsim}{\mbox{\raisebox{-.3em}{$\stackrel{<}{\sim}$}}}
\def \lleq {\lower0.9ex\hbox{ $\buildrel < \over \sim$} ~}

\title{Perfect fluid cosmologies with varying light speed }

\author{Luis P. Chimento and Alejandro S. Jakubi      \\
{\it Departamento de F\'{\i}sica,  }\\
{\it  Facultad de Ciencias Exactas y Naturales, }\\
{\it Universidad de Buenos Aires }\\
{\it  Ciudad  Universitaria,  Pabell\'{o}n  I, }\\
{\it 1428 Buenos Aires, Argentina.}}

\maketitle
\date{}

\begin{abstract}

We have found exact constant solutions for the cosmological density parameter
using a generalization of general relativity that incorporates a cosmic
time-variation of the velocity of light in vacuum and the Newtonian
gravitation constant. We have determined the conditions when these solutions
are attractors for an expanding universe and solved the problems of
the Standard Big Bang model for perfect fluids.

\end{abstract}


\newpage

\section{Introduction}

Recent observations  suggest a universe that is lightweight: matter density
about one-third the critical value, is accelerating, and is flat. The
acceleration implies the existence  of  cosmic dark energy that overcomes the
gravitational self-attraction of matter and causes the expansion to speed up.
This has once more drawn attention to the possible existence, at the present
epoch, of a small positive cosmological term. Determination of the Hubble
constant also seem to point in the same direction \cite{fhp}.

Albrecht and Magueijo \cite{alb} have investigated possible
cosmological consequences of a time variation in the velocity of light in
vacuum. In particular it offers new ways of resolving the problems of the
standard Big-Bang cosmology, distinct from their resolutions in the context of
the inflationary paradigm \cite{infl} or the pre-Big-Bang scenario of
low-energy string theory \cite{prebb}. Moreover, in contrast to the case of
the inflationary universe, varying $c$ may provide an explanation for the
relative smallness of the cosmological constant today.
An action formalism describing the speed of light as a minimally coupled
scalar is given in Ref. \cite{Barrow}. We will follow this formalism.
Several other theories including a variable speed of light have been proposed
recently \cite{Clayton99} \cite{Moffat} \cite{Drumond}.

In refs.\cite{alb} and \cite{Barrow} it was considered a model containing a
fluid whose equation of state obeys a linear baryotropic law with constant
adiabatic index. The study presented in ref.\cite{alb} envisages a sudden fall
in the speed of light, precipitated by a phase transition or some shift in the
values of the fundamental constants, and explores the general consequences
that might follow from a sufficiently large change. On the other hand in
\cite{Barrow} it was proposed that the speed of light has a power-law
dependence on the scale factor.

In this paper we investigate a varying speed of light scenario for a perfect
fluid without restriction on the equation of state. We identify the attractor
solutions and use them to perform the calculations. Thus we show that  the
cosmological field equations can be solved in general, defining the law of
variation for $c(t)$. This allows us to give simple solutions to the flatness,
coincidence, quasiflatness and horizon problems \cite{Peebles} \cite{CMBdata}
\cite{Bar2}, even when the strong energy condition is satisfied.

\section{The variable $c$ model}

In \cite{alb} it was argued that a time-variable $c$ should not introduce
changes in the curvature of the space-time  in the cosmological frame and that
Einstein's equations must still hold. They have assumed that the universe is
spatially homogeneous and isotropic, so that there are no spatial variations
in $c$ or $G$. This leads to the requirement that the Lemaitre equations still
retain their form with $c(t)$ and $G(t)$ varying. Thus the expansion scale
factor obeys the equations

\begin{eqnarray}
\frac{\dot a^2}{a^2} &=&\frac{8\pi G\rho }3-\frac{Kc^2(t)\ }{a^2}
\label{fr} \\
\ddot a &=&-\frac{4\pi G}3\left[\rho +\frac{3p}{c^2(t)}\right]a  \label{ac}
\end{eqnarray}

\noindent where $p$ and $\rho $ are the density and pressure of the perfect
fluid, and $K$ is the metric curvature parameter. This perfect fluid may
comprise several components like clustered matter with energy density $\rho_m$
and pressure $p_m$, and a cosmological term that may be described as a perfect
fluid with a stress satisfying $ p_\Lambda=-\rho_\Lambda c^2$ where

\begin{equation} \label{rholambda}
\rho_\Lambda=\frac{\Lambda c^2}{8\pi G}
\end{equation}

\noindent

From
(\ref{fr}) and (\ref{ac}), we find the generalized conservation equation
incorporating possible time variations in $ c(t)$ and $G(t)$,

\begin{equation}
\dot \rho +3\frac{\dot a}a(\rho +\frac p{c^2})=\frac{
3Kc\dot c}{4\pi Ga^2}-\rho\frac{\dot G}{G}.  \label{cons}
\end{equation}

\noindent The cosmological density parameter $\Omega$ is defined as the ratio
of as the
density of the universe with the critical density $\rho _c =(3/8\pi G)H^{2}$,
where $H$ is the Hubble variable. The critical density defines the $K=0$
solution of Eq. (\ref{fr}). The fractional contributions to the right-hand
side of Eq. (\ref{fr}), are given by $\Omega_m \equiv \rho_m /\rho_c$,
$\Omega_{\Lambda} \equiv \rho_\Lambda /\rho_c = \Lambda c^2/(3 H^2)$, and
$\Omega_K \equiv -Kc^2/(aH)^2$, respectively. Thus, inserting

\begin{equation}
\Omega \equiv \frac \rho {\rho _c}=\frac{8\pi G\rho }{3H^2}  \label{om1}
\end{equation}

\noindent
in (\ref{fr}), we have

\begin{equation}
{\frac \Omega {\Omega -1}}={\ }\frac{8\pi G\rho a^2 }{3Kc^2(t)}.\
\label{om2}
\end{equation}

\noindent Differentiating it, and using the conservation equation (\ref{cons})
we find the dynamical equation for the cosmological density parameter

\begin{equation} \label{dOmega2}
\dot\Omega=\left(\Omega-1\right)\left[
\left(1+\frac{3p}{\rho c^2}\right)\Omega H+\frac{2\dot c}{c}\right]
\equiv f\left(\Omega\right)
\end{equation}

\noindent It is important to obtain exact solutions of Eq. (\ref{dOmega2}) in
order to evaluate the effects of a perfect fluid, varying $G$ and $c$ on the
expansion dynamics. In particular, as we shall see bellow, stable constant
solutions of this equation are relevant to solve several cosmological
conundrums like the flatness, coincidence, quasi-flatness, and horizon
problems.
Eq. (\ref{dOmega2}) has three constant solutions: $\Omega=0$, when $\dot c=0$,
that corresponds to the Milne universe, $\Omega=1$ that represents the flat
universe and $\Omega=\Omega_0$ that arises when the squared bracket in
(\ref{dOmega2}) vanishes. To show that these solutions are dynamical
attractors we will use the fact that a constant solution $f(\Omega_s)=0$ of
(\ref{dOmega2}) is asymptotically stable provided $f'\left(\Omega_s\right)<0$.

\section{The Flatness Problem}

The combined measurements of the cosmic microwave background  temperature
fluctuations and the distribution of galaxies on large scales began to suggest
that the universe may  be flat~\cite{OS95}. Within standard General
Relativity, where $\dot G\equiv 0\equiv \dot c$, the conservation equation
(\ref {cons}) gives

\begin{equation} \label{rhoRG}
\rho=\rho_0 \exp\left[-3\int dt \left(1+\frac{p}{\rho c^2}\right)H\right]
\end{equation}

\noindent Hence the curvature term will dominate the matter density term at
large $a$ whenever the matter stress obeys the strong energy condition (SEC)
$\rho +3p/c^2\ge 0$. Equivalently, Eq. (\ref{dOmega2}) shows the solution
$\Omega=1$ is unstable so that extreme fine tuning of the conditions at the
early universe seems to be required to accommodate the observations. This is
called the flatness problem \cite{Peebles}. The solution postulated by
inflation requires a sufficiently long period of evolution in the early
universe during which the expansion was dominated by a gravitationally
repulsive stress that violated SEC. In this way the evolution would have been
driven very close to $\Omega=1$. Also Eq. (\ref{ac}) implies a period with
superluminal expansion as $\ddot a>0$.

From Eq. (\ref{dOmega2}) we see that a decreasing speed of light ($\dot
c/c<0$) would also drive $\Omega$ to 1. In fact, when the speed of light
changes with $\dot c/c <-\left(1+3p/\rho c^2\right)H/2$, the solution
$\Omega=1$ is asymptotically stable. Thus in a varying speed of light (VSL)
scenario the flatness problem can be solved even when SEC is satisfied. No
fine-tuning of the initial conditions is needed in this scenario and no
further specification of the equation of state is required.

\section{The Coincidence Problem}

In the last years there has been a renewed interest in the possibility that a
positive  cosmological constant may dominate the total energy density in the
universe. Interest in the cosmological constant stems from several directions.
Dynamical estimates of the amount of clustered matter yield a conservative
upper limit $\o_m \lleq 0.3$, and recent observations of Type 1a supernovae
indicate an accelerating universe. Combining them with the observations
suggesting $\Omega_K\simeq 0$ we conclude that $\o_m + \o_\l \simeq 1$, where
the density ratio parameter
$\epsilon=\rho_m/\rho_\Lambda=\Omega_m/\Omega_\Lambda$ has the current value
$\epsilon_0\simeq 3/7$ \cite{Bahc}.

The puzzle with the cosmological term is explaining why $\rho_\Lambda$ and the
matter energy density $\rho_m$ should be comparable today. Throughout the
history of the universe the two densities decrease at different rates and so
it appears that the conditions in the early universe have to be set very
carefully in order for the energy densities to be comparable today.  We refer
to this issue of initial conditions as the coincidence problem \cite{CMBdata}.

From these observational evidences, it is natural to take a step beyond
Einstein's original hypothesis and consider that the $\l$-term is not a
constant, but rather, describes a new dynamical degree of freedom.
Neither observational data, nor inflationary considerations tell us that a
cosmological term is constant. Between different mechanisms explored
in the literature to solve this problem the VSL theory provides us a simple
way to obtain a decaying effective cosmological term $\Lambda c(t)^2$.
Following \cite{alb} we adopt this model to find a stable solution of the
Einstein equations with constant $\epsilon$.

Integrating (\ref{cons}) for $\Omega=1$ we get

\begin{equation} \label{rho1}
\rho(t)=\frac{\rho_0}{G} \exp\left[-3\int
dt\,\left(1+\frac{p}{\rho c^2}\right) H\right]
\end{equation}

\noindent
where $\rho=\rho_m+\rho_\Lambda$. Hence, using (\ref{rholambda})
we find that $\epsilon$ has the constant solution

\begin{equation} \label{epsilon0}
\epsilon_0=\frac{8\pi\rho_0}{\Lambda c_0^2}-1
\end{equation}

\noindent
provided that the speed of light decreases as

\begin{equation} \label{c2}
c(t)=c_0\exp\left[-\frac{3}{2}\int
dt\,\left(1+\frac{p}{\rho c^2}\right) H\right]
\end{equation}

\noindent where $c_0$ is a positive constant (we assume the dominant energy
condition throughout). From (\ref{c2}) and (\ref{rho1})
we have

\begin{equation}
\label{r}
\rho=\frac{\rho_0 c^2}{c_0G}
\end{equation}

\noindent This solution is an attractor as in the neighbourhood of $\Omega=1$
we get $\dot\Omega\simeq -2H\left(\Omega-1\right)$. In this way we have solved
the coincidence problem in the framework of the VSL scenario as $\epsilon$
approaches the attractor solution (\ref{epsilon0}) without a fine-tuning of
the initial conditions and without restriction on the equation of state. In
this asymptotic regime, assuming cold dark matter and using (\ref{c2}), we
find that $c\simeq a^{-3\epsilon_0/[2(1+\epsilon_0)]}$.

\section{Quasiflatness problem}

In general it is accepted that our universe has a low-mass-density
($\Omega_m<1$). The determination of the universe's mass density is  currently the
best-studied cosmological parameter and its low value is indicated by a number
of independent methods for the study of clusters of galaxies. They include the
mass-to-light ratio, the baryon fraction, the cluster abundance and the mass
power spectrum \cite{Bahc}. Thus, if the energy density of our universe were
dominated by clustered matter we would find a problem related to the flatness
problem: a low $\Omega$ universe with $\Omega\simeq O(1)$ also requires
extreme fine tuning of initial conditions. This is the quasi-flatness problem
and we shall see that it also has solution within the VSL framework.

The constraint associated with the constant solution $\Omega_0$ determines a
first order differential equation for $c(t)$ whose general solution
becomes

\begin{equation} \label{ct0}
c(t)=c_0 a^{\Omega_0} \exp\left[-\frac{3\Omega_0}{2}\int
dt\,\left(1+\frac{p}{\rho c^2}\right)  H\right]
\end{equation}

\noindent We note that this expression leads to a decreasing speed of light
provided that SEC holds. We can solve  (\ref{cons}) using
(\ref{ct0})

\begin{equation} \label{rhot}
\rho(t)=\rho_0 \frac{a^{2\left(\Omega_0-1\right)}}{G}
\exp\left[-3\Omega_0\int
dt\,\left(1+\frac{p}{\rho c^2}\right) H\right]
\end{equation}

\noindent
hence we find

\begin{equation} \label{rhoc}
\rho=\frac{\rho_0}{c_0^2}\frac{c^2}{a^2G}
\end{equation}

\noindent Using (\ref{fr}), (\ref{om1}) and (\ref{rhoc}), we get this
a relationship between the cosmological density parameter and the integration
constants

\begin{equation} \label{omega0}
\Omega_0=\frac{1}{1-\frac{3Kc^2_0}{8\pi\rho_0}}
\end{equation}

\noindent Integrating (\ref{fr}) for the energy density (\ref{rhoc}), we
express the scale factor as

\begin{equation} \label{at0}
a(t)=\sqrt{\frac{K}{\Omega_0-1}}\int c(t)\,dt
\end{equation}

Using (\ref{ct0}), in a neighbourhood of the solution (\ref{omega0}),
Eq. (\ref{dOmega2}) becomes

\begin{equation} \label{domega4}
\dot\Omega\simeq \left(\Omega_0-1\right)\left(1+3\frac{p}{\rho c^2}\right)H
\left(\Omega-\Omega_0\right)
\end{equation}

\noindent Hence the solution $\Omega_0<1$ is an attractor provided SEC
holds, solving the quasi-flatness problem for any perfect fluid that satisfies
this condition.

In the particular case that we choose the linear baryotropic equation of state
$p=\left(\gamma-1\right)\rho c^2(t)$ with constant adiabatic index $\gamma$,
we obtain from (\ref{ct0})

\begin{equation} \label{ctb}
c(t)=c_0 a^n
\end{equation}

\begin{equation} \label{Omega0b}
\Omega_0=\frac{2n}{2-3\gamma}
\end{equation}

\noindent In this case SEC holds when $\gamma>2/3$ that means $n<0$ when
$\Omega_0>0$. From (\ref{at0}) the scale factor is

\begin{equation} \label{atb}
a(t)\propto \Delta t^{\frac{1}{1-n}}
\end{equation}

\noindent
In this way our exact solution is the late time limit of the solution found in
\cite{Bar2}.

\section{The Horizon Problem}

One of the most puzzling features of the Standard Big Bang model is the
presence of cosmological horizons. At any given time any observer can only see
a finite region of the Universe. Since the horizon size increases with time we
can now observe many regions in our past light cone which are causally
disconnected. The fact that these regions have the same properties is puzzling
as they have not been in physical contact \cite{infl}\cite{Peebles}. This
would have required an extraordinary fine-tuning of the initial conditions in
the early Universe. Usually  this problem is solved in the inflationary
scenario again appealing to sufficiently long period violating SEC. We will
show that a  VSL scenario may also solve this problem without its violation.

The observed Universe was already smooth at the nucleosynthesis era ($\simeq
1$s). Hence the horizon problem has to be solved at an earlier time, $t_1$
say. The size of the connected region at time $t_1$ with events at an initial
time $t_0$ is proportional to the integral

\begin{equation} \label{xh}
\int_{t_0}^{t_1} dt' \frac{c(t')}{a(t')}
\end{equation}

\noindent 
and the particle horizon corresponds to the limit $t_0\to 0$.
When the curvature becomes negligible the scale factor for the scenario
solving the coincidence problem is given by

\begin{equation} \label{ath}
a(t)=a_0\exp\left[\sqrt{\frac{8\pi\rho_0}{3}}
\int dt\, \frac{c}{c_0}\right]
\end{equation}

\noindent In this case the integral (\ref{xh}) diverges in the limit $t_0\to
0$. The same occurs when this integral is evaluated using the expressions for
the speed of light (\ref{ct0}) and scale factor (\ref{at0}) of the scenario
solving the quasiflatness problem. Hence in the VSL cosmology the size of the
connected region at $t_1$ can be made arbitrary larger than the size of our
past light cone at $t_1$ by choosing the initial time $t_0$ small enough.
This solves the horizon problem without violation of SEC.

\section{Discussion}

In this letter we have considered a varying-$c$ theory recently proposed by
Albrecht and Magueijo \cite{alb} as a new way of solving the flatness,
coincidence, quasiflatness and horizon problems for any perfect fluid.

We have found the constant solutions for the cosmological density parameter of
the Einstein equations and the conditions that determine when these solutions
are attractors for an expanding universe. In this way we have solved several
problems of the Standard Big Bang cosmology without requiring a stage when the
strong energy condition is violated. No specific form for the equation of
state of the perfect fluid has been imposed, and in this way a wide range of
matter sources can be accommodated within our model. Also we have proposed a
better way to understand a small but nonzero cosmological constant, as
indicated by a number of recent observational studies.

It is interesting to mention that a time dependent $G$ cannot solve the
flatness, coincidence or quasiflatness problems as it does not appear in Eq.
(\ref{dOmega2}). We have not touched in this letter the issue of entropy
production within the VSL scenario. This will be the subject of a future
paper.

\textbf{Acknowledgements}

This research has been supported by the University of Buenos Aires under
project TX-93. We are grateful to the University of the Basque Country for
partial support under project UPV 172.310-G02/99.

\end{document}